\documentclass[fleqn,twoside,psfig]{article}
\usepackage{espcrc2}

\usepackage{psfig}

\newcommand{\bei}{\begin{itemize}}
\newcommand{\eei}{\end{itemize}}
\newcommand{\beq}{\begin{equation}}
\newcommand{\eeq}{\end{equation}}
\newcommand{\beqn}{\begin{eqnarray}}
\newcommand{\eeqn}{\end{eqnarray}}
\newcommand{\beqns}{\begin{eqnarray*}}
\newcommand{\eeqns}{\end{eqnarray*}}

\RequirePackage{xspace}
\usepackage{relsize}
\def\babar{\mbox{\slshape B\kern-0.1em{\smaller A}\kern-0.1em
    B\kern-0.1em{\smaller A\kern-0.2em R}}}
\def\PL{{\it Phys. Lett.}}

\def\ea{{\it et al.}}
\def\ee{$e^+e^-$}

\def\ee{$e^+e^-$}

\def\rar{\rightarrow}
\title{Progress on R Measurement through ISR with \babar} 

\author{Michel Davier\\
	Laboratoire de l'Acc\'el\'erateur Lin\'eaire\\
        IN2P3/CNRS et Universit\'e de Paris-Sud\\
        91898 Orsay, France\\
	E-mail: davier@lal.in2p3.fr}

\begin{document}

\begin{abstract}

\vspace{1pc}
The status of the ISR analysis with \babar\ is presented. Preliminary 
results are presented on the process $e^+e^- \rightarrow 2\pi^+ 2\pi^-$
and J/$\psi$ decays including a determination of its total width and
the branching ratios into $2\pi^+ 2\pi^-$, $K^+K^-\pi^+\pi^-$ and
$2K^+2K^-$ modes. 
\end{abstract}

\maketitle
\setcounter{page}{1}
%
%
\section{Introduction}
\label{sec_introduction}

High-luminosity \ee\ storage rings designed for the study of CP violation,
such as KEK-B, PEP-B, and DA$\Phi$NE, have
opened a new possibility to measure \ee\ annihilation using initial-state
radiation (ISR)~\cite{arbuzov,binner,benayoun}. A large mass range becomes
accessible in a single experiment, especially at B factories, contrary to 
the case with fixed energy machines which are optimized only in a limited
region. The broad-band coverage may result into a better understanding of 
systematic effects since only one experimental set-up is involved.

Apart from their own physics interest, ranging from hadronic spectroscopy
to tests of QCD, measurements of the ratio 
$R=\sigma (e^+e^- \rar {\rm hadrons})/ \sigma _{\rm pt}$
provide the necessary input to dispersion integrals for
computing hadronic vacuum polarization contributions. This is indeed the
case when calculating the theoretical prediction for the muon magnetic anomaly
$a_\mu$ and the running of the electromagnetic coupling to the Z mass,
$\alpha (M_{\rm Z}^2)$, one of several ingredients in precision tests of
the electroweak theory bearing on such fundamental issues as the structure
of the Higgs sector.

The cross section for the ISR process $e^+e^- \rightarrow f \gamma$ 
with a particular final state $f$ is related to the cross section 
$\sigma _f (s)$ for the direct annihilation $e^+e^- \rightarrow f$ 
through
\beq
   \frac {d \sigma (s,x)}{d x}~=~W(s,x)~\sigma_f [s(1-x)]
\eeq
where $x=2E_\gamma^*/\sqrt s$, $E_\gamma^*$ being the radiated photon 
energy in the \ee\ frame and $\sqrt s$ the nominal centre-of-mass 
energy of the collider. The quantity $s'=s(1-x)$ represents the mass 
squared of the final-state $f$ system.
The radiator function $W(s,x)$ providing the virtual photon spectrum
can be computed including radiative corrections to an accuracy better
than $1\%$~\cite{kuehn_pisa}. The production of ISR photons is strongly 
peaked along the initial beams, but for $\sqrt s \sim 10$~GeV the fraction
at large angle is relatively large, {\it i.e.} $10-15\%$ in the \babar\ 
central detector.

The measurement of the corresponding leptonic process 
$e^+e^- \rar \mu^+ \mu^- \gamma$ provides the ISR luminosity. Thus the
Born cross section $\sigma_f (s')$ is obtained through
\beq
 \sigma_f (s')~=~\frac {\Delta N_{f \gamma}~\epsilon_{\mu\mu}~
   (1+\delta_{FSR}^{\mu\mu})}{\Delta N_{\mu\mu \gamma}~\epsilon_f~
   (1+\delta_{FSR}^f)}~\sigma_{\mu\mu} (s')
\label{isr_ratio}
\eeq
where $\Delta N_{f \gamma}$ is the number of detected $f \gamma$ events 
in the bin of width $\Delta s'$ centred at $s'$,
$\epsilon_f$ is the detection efficiency for the final state $f$ and
$\delta_{FSR}^f$ is the fraction of the cross section for 
$e^+e^- \rar f \gamma$ where the photon is emitted by the final-state
particles. The latter quantity can be sizeable for the $\mu\mu$ channel, but
negligible for most of the low-energy hadronic states which have vanishingly
small cross sections at the nominal energy $\sqrt s$.
The radiative corrections for the initial state and vacuum
polarization effects cancel in the ratio~(\ref{isr_ratio}), as does the
photon efficiency. 

The \babar\ detector at the PEP-II asymmetric \ee\ storage ring 
is well suited for this study. The different triggers routinely 
used allow one to select ISR processes with a $95-98\%$ efficiency.
A complete description of \babar\ can be found elsewhere~\cite{babar_det}.
All main detector components are used in this analysis as particle 
identification plays a crucial role in separating the different channels.
It is also important in correctly computing the $s'$ value 
for each event as it is obtained from the produced particles momenta 
since the photon energy is rather insensitive to $s'$ in the
low-energy regime of interest. In particular, K-$\pi$ separation in quartz
radiators (DIRC) is an essential ingredient. The information from the tracking
system (silicon vertex detector and drift chamber) is used to measure the
angles and the momentum of the charged particles. Muons are identified in the
instrumented flux return (IFR) of the superconducting solenoidal magnet,
while photons are detected in the fine-grain CsI calorimeter.

The data used in this analysis correspond to an integrated luminosity 
of 89.3 fb$^{-1}$, collected both at the $\Upsilon(4S)$ and in the nearby 
continuum.

\section{Event preselection}

ISR events are preselected requiring a large-angle photon in the central part
of the detector with an energy in the centre-of-mass larger than 3 GeV. This
requirement is loosened down to 0.5 GeV for 2-track events which have at least one 
track identified as a muon, in order to collect a large statistics of muon tracks
for calibrating the muon-identification performance. The number of tracks pointing
well to the interaction region is required to be even and larger than two, and their total 
charge should be equal to zero. A loose kinematic constraint is applied by asking 
the ISR photon candidate to be inside a 0.3 radian cone whose axis is along 
the missing momentum vector constructed from all the recorded recoiling particles.

Essentially all low-energy exclusive ISR final states are retained by these general
selection criteria, while about $2\%$ of all \babar\ events are kept.

\section{The di-muon final state}

The process $e^+e^- \rar \mu^+ \mu^- \gamma$ plays an important role in this
analysis. For $\mu\mu$ masses less than a few GeV, it provides the measurement
of the ISR luminosity, while for masses larger than 2 GeV, where hadron pairs are 
negligible, it enables one to  measure the muon identification efficiency.
For this latter purpose, a muon candidate is tagged by the opposite identified
muon. The large statistics of muons collected in this way efficiently covers the
acceptance of the detector so that fine-grained efficiency maps in momentum, polar
and azimutal angles can be constructed.

A 1C kinematic fit using the two muon tracks effectively removes background from
$\tau^+ \tau^-$ events. After the $\mu\mu$ mass and $\chi^2$ cuts, the purity of
the muon sample is found to be better than 0.999. In addition, the fit improves the $\mu\mu$ 
mass resolution which decreases from 16 MeV down to 8 MeV at the J/$\psi$ mass.

\begin{figure}[t]
\centerline{\psfig{file=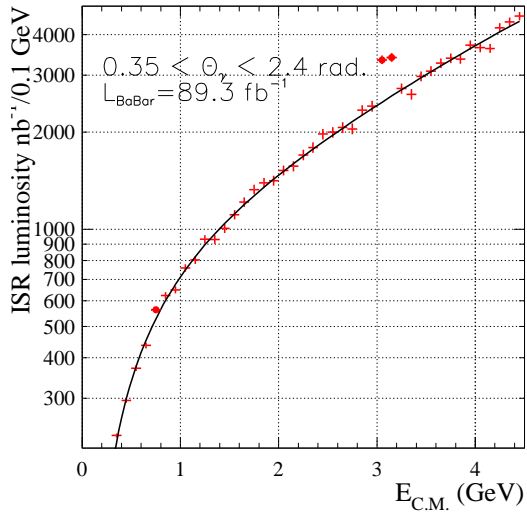,width=70mm}}
\caption[.]{\it The differential ISR luminosity in 100-MeV bins obtained from
           the measurement of $e^+e^- \rightarrow \mu^+ \mu^- \gamma$. The
           points near the $\rho(770)$ and J/$\psi$ resonances are excluded,
           because of pion misidentification as muon in the first case and
           the non-QED nature of the second process.}
\label{lumi}
\end{figure}

Fig.~\ref{lumi} shows the derived spectrum of ISR luminosity per 100 MeV energy bins.
Thus, the \babar\ luminosity of 89.3 fb$^{-1}$ corresponds to an $e^+e^-$ energy 
scan taken all together with 0.7 pb$^{-1}$/100 MeV at 1 GeV and 3.6 pb$^{-1}$/100 MeV
at 4 GeV. It provides a statistically competitive hadron sample, especially in the
range between 1.4 and 3 GeV.

\section{Hadronic channels}

The main approach to the R measurement is the study of exclusive hadronic channels,
constrained by a kinematic fit and with identified charged particles. Work is in
progress along these lines, including the measurement of the following final states:
$\pi^+ \pi^-$, $K^+ K^-$, $p \overline {p}$, $\pi^+ \pi^- \pi^0$, $4\pi$, $5\pi$,
$6\pi$, $\pi\pi\eta$, $K \overline {K}\pi$, $K \overline {K}\pi\pi$,
$2K 2\overline {K}$, $K \overline {K}\eta$, with charged and 
neutral ($K^0_s$) kaons. Another approach, based on an inclusive
method where the hadronic mass is deduced from the ISR photon energy, is also in progress,
with the difficulty that the resolution deteriorates rapidly for low recoil masses.

The study of the $\pi^+ \pi^-$ final state is an important part of the programme,
especially in view of persisting discrepancies between existing $e^+ e^-$ results and
isospin-corrected $\tau$ data~\cite{davier_tausf,davier_g-2}. Such a study requires
a good control of systematic uncertainties at the $1\%$ level or better. The ISR
method has several advantages compared to the standard fixed-energy annihilation:
(i) a large mass range is simultaneously covered in a single experiment, (ii) the 
cross section can be measured down to threshold with excellent acceptance, and (iii)
easier particle identification can be achieved, thanks to the Lorentz boost given to the hadronic
system.

Since the $\pi^+ \pi^-$ and $\mu^+ \mu^-$ final states have the same 
topology, it is advantageous to directly measure the ratio 
$N_{\pi\pi}/N_{\mu\mu}$ since many effects cancel: the $e^+ e^-$ luminosity,
ISR and vacuum polarization radiative corrections, the photon detection efficiency.
Small corrections still need to be applied concerning trigger efficiencies 
(calorimeter triggers are more efficient for pions), tracking efficiency (because
pions can produce secondary interactions in the tracking detector), FSR corrections.
All these effects can in fact be studied directly on the data. The major
experimental task remains the determination of the particle identification efficiency matrix, 
using pure samples of known particles, as outlined above for the muons, as a function
of momentum and the 2-dimensional location of the particle in the relevant 
detectors. Such a detailed study is still in progress.

\begin{figure}[t]
\centerline{\psfig{file=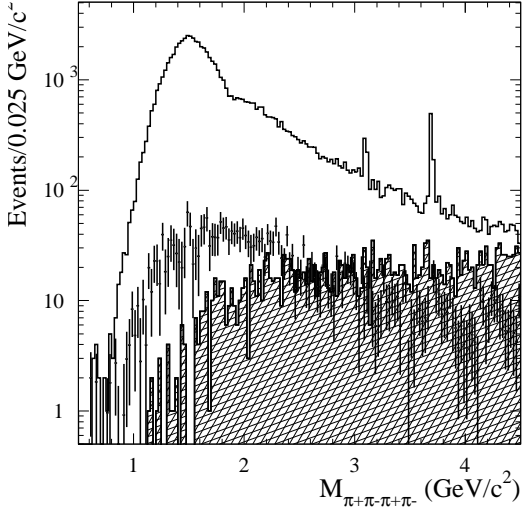,width=70mm}}
\caption[.]{\it The 4-pion invariant mass distribution for kinematically
           constrained events of the type 
           $e^+e^- \rightarrow 2\pi^+ 2\pi^- \gamma$. 
           No muon/pion separation is applied. The points with error bars
           indicate the level of background obtained from a study of the
           $\chi^2$ distribution of the kinematical fit. The cross-hatched
           histogram corresponds to the non-ISR background as given by the
           JETSET simulation of annihilation events. Sharp signals are seen
           at the J/$\psi$ and $\psi'$ masses.}
\label{4pi_plot}
\end{figure}

\begin{figure}[t]
\centerline{\psfig{file=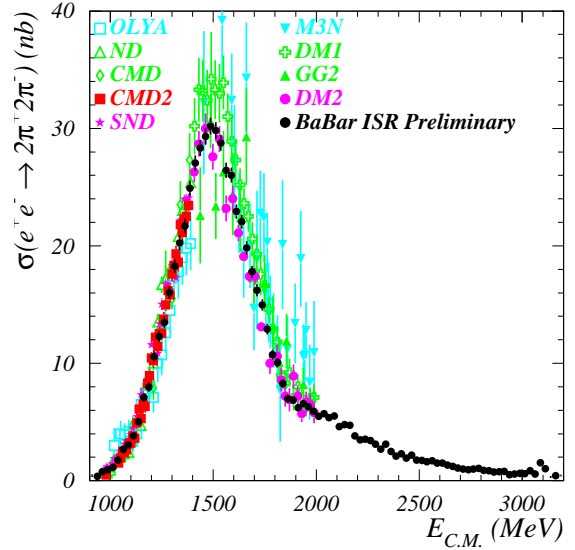,width=75mm}}
\caption[.]{\it The $e^+e^- \rightarrow 2\pi^+ 2\pi^-$ cross section obtained
            from ISR at \babar\ in comparison with all existing \ee\ data.}
\label{4pi_xsection}
\end{figure}

Preliminary results are available for the final state with four charged particles, 
namely $2\pi^+ 2\pi^-$, $K^+ K^- \pi^+ \pi^-$, and $2K^+ 2K^-$. Fig.~\ref{4pi_plot}
shows the mass distribution of the events satisfying the $2\pi^+ 2\pi^-$ kinematics
and K/$\pi$ identification constraints. No muon identification was applied as no 
corresponding background is expected in this four-track sample. The data are seen 
to be dominated by a broad peak at 1.5 GeV originating from the $\rho(1450)$ and 
$\rho(1700)$ resonances, a shoulder near 2 GeV, and sharp peaks at the J/$\psi$ and 
$\psi'$ masses. The latter signal is the result of the decay chain 
$\psi' \rightarrow \pi^+ \pi^-$~J/$\psi$, J/$\psi \rightarrow \mu^+ \mu^-$. The
background from other hadronic channels is small and subtracted in each mass bin.
Overall, about 70k events are obtained leading to small statistical uncertainties.
The normalization uncertainty is estimated to be $5\%$, dominated by the luminosity 
determination and the dependence of the detection efficiency on the dynamics 
(below 2 GeV the intermediate state $\pi ~{\rm a}_1$ is found to be dominant).
The derived cross section is given in Fig.~\ref{4pi_xsection}. The accuracy of the
\babar\ data is comparable to the latest most precise results from CMD-2~\cite{cmd2_4pi} 
and SND~\cite{snd_4pi} below 1.4 GeV and the agreement is good. Between 1.4 and 2 GeV 
the quality of the \babar\ data exceeds that from DCI and Adone. The range above 
2 GeV is only covered by \babar.

One can illustrate the impact of the \babar\ results on the $2\pi^+ 2\pi^-$ annihilation
cross section obtained through ISR by computing its contributions to the lowest-order
hadronic component of the muon magnetic anomaly. Using the same energy range from
threshold up to 1.8 GeV as in Ref.~\cite{dehz03} for a convenient comparison, one gets
\beq
   a_\mu^{\rm had,LO}[4\pi]~=~(12.95 \pm 0.64_{exp} \pm 0.13_{rad})~10^{-10}
\eeq 
to be compared to the values~\cite{dehz03} obtained using all previously available 
$e^+ e^-$ data, $(14.21 \pm 0.87_{exp} \pm 0.23_{rad})~10^{-10}$, and from $\tau$ data
on the difficult $\tau^- \rightarrow \pi^- 3\pi^0 \nu_\tau$ decay mode,
$(12.35 \pm 0.96_{exp} \pm 0.40_{SU(2)})~10^{-10}$. In the above results, the uncertainties
have been separated into experimental sources '$exp$' (statistics and systematics), 
radiative corrections '$rad$' and isospin-breaking corrections '$SU(2)$'. The \babar\
results lead to a significant improvement in this channel. 

\section{Radiative return to the J/$\psi$}

For a narrow state of mass $M$ such as J/$\psi$, decaying into a final state $f$, one has
\beq
   \sigma_{{\rm J/}\psi}~=~ \frac {12 \pi^2 \Gamma_{ee} B_f} {M s} W(s,x_0)
\eeq
with $x_0=1-M^2/s$. The total J/$\psi$ ISR cross section amounts to 0.036 nb for
$s=(10.58)^2$~GeV$^2$, leading to $4~10^6$ events for the present 130 fb$^{-1}$ of
\babar\ luminosity (the angular acceptance of the detected ISR photon is about 0.1).
 
\begin{figure}[t]
\centerline{\psfig{file=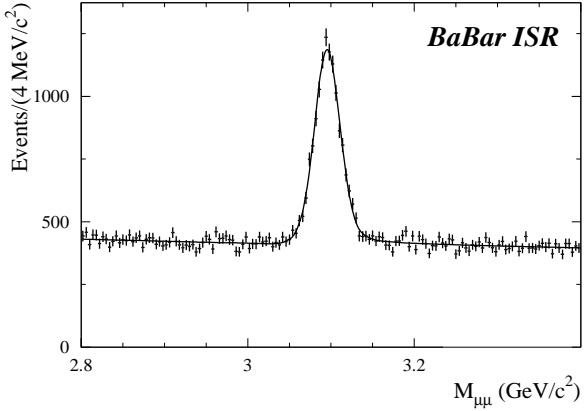,width=80mm}}
\caption[.]{\it The $\mu^+\mu^-$ invariant mass distribution in the vicinity
            of the J/$\psi$ resonance.}
\label{psitomumu}
\end{figure}

For the $\mu^+ \mu^-$ final state, one can directly compare the rate in the J/$\psi$
peak (about 8000 events) to the continuum (see Fig.~\ref{psitomumu}) and derive 
the value of the product
\beqn
   \Gamma_{{\rm J/}\psi \rightarrow e^+ e^-} B_{{\rm J/}\psi \rightarrow \mu^+ \mu^-}~&=&~  \nonumber \\
      (0.330 \pm 0.008_{stat} \pm 0.007_{syst})~{\rm keV} 
\eeqn
The systematic error receives contributions from the background estimation, the J/$\psi$ 
line shape, the radiative corrections and Monte Carlo statistics. Using the world
averages for $B_{{\rm J/}\psi \rightarrow \mu^+ \mu^-}$ and 
$B_{{\rm J/}\psi \rightarrow e^+ e^-}$, one can derive the J/$\psi$ electronic and total
widths, $\Gamma_{{\rm J/}\psi \rightarrow e^+ e^-}=(5.61 \pm 0.20)$~keV and
$\Gamma_{{\rm J/}\psi}=(94.7 \pm 4.4)$~keV, to be compared to the world average 
values~\cite{pdg} of $(5.26 \pm 0.37)$~keV and $(87 \pm 5)$~keV, respectively.

From the mass spectrum given in Fig.~\ref{4pi_plot}, the following 
quantities can be derived:
\beqn
    \Gamma_{{\rm J/}\psi \rightarrow e^+ e^-} B_{{\rm J/}\psi \rightarrow 4\pi}~&=&~ \nonumber \\
      (1.95 \pm 0.14_{stat} \pm 0.13_{syst})~10^{-2}~{\rm keV} \\ 
   \Gamma_{\psi' \rightarrow e^+ e^-} B_{{\rm J/}\psi \rightarrow \mu^+ \mu^-}
        B_{\psi' \rightarrow {\rm J/}\psi \pi^+ \pi^-}~&=&~ \nonumber \\
      (4.50 \pm 0.18_{stat} \pm 0.22_{syst})~10^{-2}~{\rm keV} 
\eeqn
Using the world averages~\cite{pdg} for the J/$\psi$ and $\psi'$ electronic widths, 
the results $B_{{\rm J/}\psi \rightarrow 4\pi}=(3.70 \pm 0.27 \pm 0.36)~10^{-3}$ and
$B_{\psi' \rightarrow {\rm J/}\psi \pi^+ \pi^-}=0.361 \pm 0.015 \pm 0.037$ are
obtained, in agreement with the world average values~\cite{pdg} of 
$(4.0 \pm 1.0)~10^{-3}$ and $0.305 \pm 0.016$, respectively.

\begin{figure}[p]
\centerline{\psfig{file=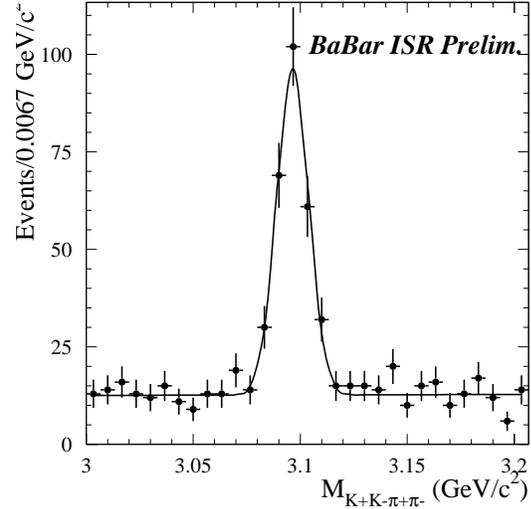,width=70mm}}
\caption[.]{\it The $K^+K^-\pi^+\pi^-$ mass distribution for ISR-produced
            $e^+e^- \rightarrow K^+K^-\pi^+\pi^-$ in the J/$\psi$ region.}
\label{2k2pi}
\end{figure}

\begin{figure}[p]
\centerline{\psfig{file=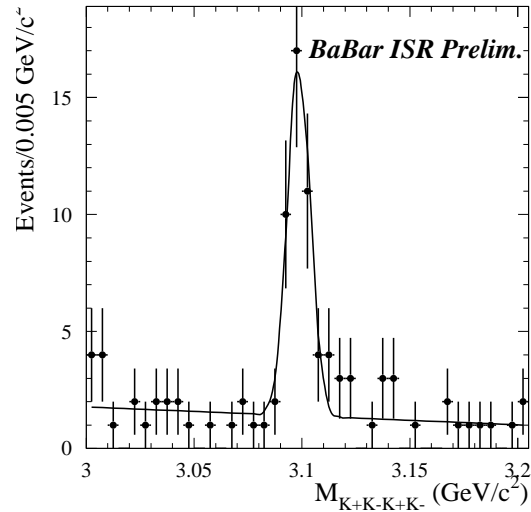,width=70mm}}
\caption[.]{\it The $K^+K^-K^+K^-$ mass distribution for ISR-produced
            $e^+e^- \rightarrow K^+K^-K^+K^-$ in the J/$\psi$ region.}
\label{4k}
\end{figure}

Taking advantage of the K/$\pi$ Cherenkov identification, one can measure the
J/$\psi$ decay rate into final states involving $K^+K^-$ pairs. The
corresponding mass spectra are given in Figs.~\ref{2k2pi} and \ref{4k}. Using
again the world average value for the electronic width, the following
branching ratios are measured: 
$B_{{\rm J/}\psi \rightarrow K^+K^-\pi^+\pi^-}=(6.25 \pm 0.50 \pm 0.62)~10^{-3}$ 
and $B_{{\rm J/}\psi \rightarrow K^+K^-K^+K^-}=(6.9 \pm 1.2 \pm 1.1)~10^{-4}$.
Both determinations are more precise than the current world averages~\cite{pdg}
of $(7.2 \pm 2.3)~10^{-3}$ and $(7. \pm 3.)~10^{-4}$, respectively.

\section{Conclusions}

Preliminary results from \babar\ on final states produced through ISR
show the high physics potential of this sample which should yield precise
measurements of \ee\ annihilation cross sections. The ISR method will allow
one to cross check the results of the 'direct' experiments, with the
significant advantage that the important energy range from threshold to
3-4 GeV can be covered in a single experiment, thereby reducing
systematic uncertainties.

The ratio R will be measured from the sum of exclusive channels, providing
input for vacuum polarization calculations, QCD studies and investigations in
hadronic spectroscopy. Amazingly, the current data is already providing
very competitive results for J/$\psi$ and $\psi'$ decays.

\section*{Acknowledgments}

I would like to thank my \babar\ colleagues for the fruitful collaboration
and congratulate Marco Incagli and
Graziano Venanzoni for organizing a very stimulating workshop.

\end{document}